\documentclass{llncs}
\usepackage{graphicx, amsmath, subfigure, cite}
\begin{document}

\title{Distributed Robust Geocast}
\subtitle{Multicast Routing for Inter-Vehicle Communication}

\author{Harshvardhan P. Joshi\inst{1} \and Mihail L. Sichitiu\inst{1} \and Maria Kihl\inst{2}}

\institute{Department of Electrical and Computer Engineering, North
Carolina State University, Raleigh NC 27606, USA. \email{hpjoshi,
mlsichit@ncsu.edu} \and Department of Communication Systems, Lund
University, SE-221 00 Lund, Sweden, \email{maria@telecom.lth.se}}

\maketitle

\begin{abstract}

Numerous protocols for geocast have been proposed in literature. It
has been shown that explicit route setup approaches perform poorly
with VANETs due to limited route lifetime and frequent network
fragmentation. The broadcast based approaches have considerable
redundancy and add significantly to the overhead of the protocol. A
completely distributed and robust geocast approach is presented in
this paper, that is resilient to frequent topology changes and
network fragmentation. A distance-based backoff algorithm is used to
reduce the number of hops and a novel mechanism to reduce redundant
broadcasts is introduced. The performance of the proposed protocol
is evaluated for various scenarios and compared with simple flooding
and a protocol based on explicit route setup.
\end{abstract}

\section{Introduction}\label{sec:Introduction}
Considerable research effort is concentrated on inter vehicle
communication (IVC), which is expected to make road transport safer
and more comfortable, while reducing travel time. The term
vehicular ad hoc networks (VANETs) is frequently used for
mobile ad hoc networks (MANETs) in the context of IVC, highlighting
its distinct characteristics. Differences include, high node velocity,
constrained mobility (to roads), anonymity of the users, and
availability of position information through global positioning
system (GPS). The high node mobility in a VANET causes frequent
changes in the network topology and the network is subject to
frequent fragmentation. Furthermore, the route lifetime drops to
nearly route discovery time for more than a few
hops~\cite{Blum:PerfVANET}. Given these characteristics, it is
generally accepted that VANETs should use some form of geographical
routing. Several algorithms that use location information for
efficient route discovery have been
proposed~\cite{MauPositionBased}. For many of the applications of
IVC listed above, especially safety and traffic control
applications, it is desirable to send a message to a particular
geographic region. The multi-cast of a message to nodes satisfying a
set of geographical criteria is called \emph{geocast}. Several
algorithms for geocasting have been proposed based on location
information \cite{MaiGeocastSurvey,JiaGeocastReview}. These
algorithms, though distributed compared to other traditional routing
protocols, require at least some state information, like the
knowledge of neighbor nodes. Keeping state information adds overhead
and consumes resources like bandwidth and memory.

In this paper we propose Distributed Robust Geocast (DRG), a geocast
approach designed for VANETs that is completely distributed, without
control overhead and state information and is resilient to frequent
topology changes. We use a distance-based backoff similar
to~\cite{LindaOvercoming,LindaDisseminating,BacIVG,BenODAM} for
directed and restricted flooding. However,
unlike~\cite{LindaOvercoming,LindaDisseminating,BacIVG,BenODAM}, our
approach is not limited to a one-dimensional road and a
one-dimensional target region. We use a state-less forwarding
algorithm that efficiently spreads the message through the target
region and ensures delivery to all relevant nodes. The forwarding
algorithm can work for two-dimensional street networks as well as
one-dimensional highways and a target region of any shape.
Furthermore, the algorithm is resilient to the underlying radio
transmission range model and can work with non-circular transmission
range models caused by fading and pathloss. The algorithm can overcome
temporary network partitioning or temporary lack of relay nodes and
has a mechanism to prevent loops. We also show a completely
distributed method for keeping a message alive in the target region
thereby ensuring that a node entering the region even after the spread
of message receives the message.

\section{Related Work}\label{sec:RelatedWork}

In a seminal work,~\cite{KoLBM} outlines two schemes for
location-based multicast to a geographical region called multicast
region. Both these schemes are based on restricted flooding and does
not require topology information. It is shown in~\cite{StoLoopFree},
that memory-less routing algorithm based on direction of
destination, such as scheme~1 in~\cite{KoLBM} or
\cite{BasagniDREAM}, do not guarantee loop-free paths. However,
routing algorithms that forward to nodes closest to the destination
or with the most forward progress are inherently loop-free.
In~\cite{StoVoronoi}, use of Voronoi diagram or convex hull for
finding a neighbor closest to the destination or having the most
forward progress is proposed. These algorithms require at least
one-hop neighbor location information, and hence introduce
additional overhead of location updates or query.

Due to its simplicity, flooding the network is a tempting solution for
IVC message dissemination.  A comparison of broadcasting techniques
for MANETs is presented in~\cite{Williams:BroadcastComparison}. The
broadcasting techniques are classified as probability based methods,
with full flooding as a specific case with broadcast probability of
100\%; area based methods, discussed below; and neighbor knowledge
methods, which require state information on 1-hop or 2-hop neighbors.
In~\cite{Ni:BroadcastStorm}, it is shown that full flooding has
significant redundancy, and can cause considerable contention and
collisions. It is proposed that, to alleviate the contention and
collision, the redundancy in broadcasts should be reduced. Some area
based schemes are proposed, which assume equal transmission range for
all nodes, and are founded on the concept that a node should not
rebroadcast unless the broadcast can significantly add to the coverage
of previous broadcasts. In the distance-based scheme, a node does not
rebroadcast unless its distance from the previous sender is above a
certain threshold. For location-based scheme a convex polygon test is
proposed to determine if a rebroadcast will add significantly to the
coverage. However, this test works only if the node has received the
same packet from at least three other nodes, a serious limitation for
a highway scenario. We propose a simple angle based test instead of
the convex polygon test which works with two nodes.

A completely distributed forwarding algorithm tailored for
inter-vehicle communication is presented
in~\cite{LindaDisseminating,LindaOvercoming}.
In~\cite{BacIVG,BenODAM}, geocast approaches based
on~\cite{LindaDisseminating,LindaOvercoming} are proposed. These
approaches are designed for collision warning application on a
one-dimensional highway scenario, and do not adapt well to a
two-dimensional city street scenario or other applications.

In ~\cite{Lochert:CityGSR,Lochert:Geo}, the problem of radio
obstacles in city affecting the performance of routing algorithms is
addressed by routing around the obstacles using greedy routing.
However, neighbor tables with frequent updates are required and
nodes need to detect street junctions.

It may be useful, or even essential, for a message to be available
to vehicles that enter the geocast region after the message has
spread. In~\cite{MaiStoredGeocast}, three different approaches to
time persistent geocast are proposed. In the server approach, the
message is stored on a central server and delivered to new nodes in
the geocast region. In the election approach, one or more nodes
within the geocast region are elected to store and deliver the
message. In neighbor approach, each node store the message and
delivers to a new neighbor either by periodic broadcasts or on
notification. In~\cite{Maihoefer:AbidingGeocast}, a numerical and
analytical evaluation of the these approaches for a random waypoint
mobility model is presented, and it is shown that approaches with
local message storage cause less network load.

\section{Distributed Robust Geocast}\label{sec:DRG}

We first define certain terms used in this and subsequent sections.
The \emph{zone of relevance (ZOR)} is the set of geographic criteria
a node must satisfy in order for the geocast message to be relevant
to that node; while, the \emph{zone of forwarding (ZOF)} is the set
of geographic criteria a node must satisfy in order to forward a
geocast message.

A \emph{coverage disk} is the disk with the transmitting node at the
center and the transmission range as the radius. All nodes within
the coverage disk receive the transmission with a probability of 1.
The \emph{coverage area} or \emph{reception area} is the area around
the transmitting node within which all the nodes are supposed to
receive fraction of transmitted packets above a threshold value. The
coverage area need not be circular, and it is a more realistic model
of radio transmission with fading, pathloss and radio obstacles.

We assume a physical model that allows for a symmetrical radio
reception, i.e., if node $A$ can receive a transmission from node $B$
with probability $x$, the reverse is also true. The symmetrical radio
model can work even in city environments, where the transmission area
is not circular but rather elongated along the streets.

\subsection{Forwarding Algorithm}
It has been shown that simple flooding causes redundant
transmissions~\cite{Ni:BroadcastStorm} resulting in significant
contention and collisions. However, the redundancy can be reduced by
selecting only those nodes with the most forward progress towards
the destination as relays. A completely distributed algorithm to
select the relay node using a backoff scheme that favors the nodes
at the edge of the transmission range was proposed
in~\cite{LindaDisseminating}. On receiving a message, each node
schedules a transmission of the message after a distance-based
backoff time. Any node that loses the backoff contention to a node
closer to the destination cancels the transmission. If each node
waits for a time inversely proportional to its distance from the
last sender before retransmitting, the farthest node will be the
first to transmit winning the contention. The distance-based backoff
can be calculated using the following formula:
\begin{equation}\label{eqn:DistBackoffSimple}
BO_{d}(R_{tx}, d) = MaxBO_{d} \cdot S_{d}
    \left(\frac{R_{tx}-d}{R_{tx}}\right),
\end{equation}
where $BO_{d}$ is the backoff time depending on the distance from
the previous transmitter, $MaxBO_d$ is the maximum backoff time
allowed, $S_d$ is the distance sensitivity factor used to fine tune
the backoff time, $R_{tx}$ is the nominal transmission range, and
$d$ is the distance of the current node from the last transmitter. A
collision avoidance mechanism like random backoff can also be added.

\subsection{Network Fragmentation}
Since VANETs are prone to frequent, though temporary, fragmentation
a mechanism to overcome them can improve the performance. One of the
approaches is periodic retransmission of the message until a new
relay transmits the message, which is treated as an \emph{implicit
acknowledgement} by the previous relay. We propose a burst of
retransmissions with short interval to overcome communication
losses, and retransmission after a long interval to overcome network
fragmentation.

A relay, after its transmission at time $t$, schedules
retransmission of the message at $t + MaxBO_d$,
using~\eqref{eqn:DistBackoffSimple}. Thus, the existing relay enters
the contention for the next transmission, but with the least
preference for winning. The minimum value for $MaxBO_d$ should be at
least the round trip time for the packet to the farthest node in the
coverage area.
\begin{align}
MaxBO_d &\geq 2 \times \textrm{(maximum end-to-end delay)}.
\label{eqn:MaxBackoffDist}
\end{align}
Selecting a value higher than this bound will result in
unnecessarily longer delays. Hence, the equality in
\eqref{eqn:MaxBackoffDist} gives the value for $MaxBO_d$. A
\emph{long backoff time} ($LongBO_d$) is used after a certain number
of retransmissions, denoted \emph{maximum retransmissions}
($MaxReTx$). A few retransmissions at short duration are needed to
make sure that the absence of implicit acknowledgement is not due to
the channel losses. However, after a few retransmissions it can be
safely assumed that an implicit acknowledgement is not received due
to network fragmentation. Hence, the next retransmission can be
scheduled after a comparatively longer period $LongBO_d$, which
allows time for the network to get repaired. The selection of value
for $LongBO_d$ is a trade-off between redundant transmissions and
end-to-end delays. The maximum value of long backoff, $MaxLongBO_d$,
should be the time it takes a vehicle to reach the relay node after
it enters the coverage area. This limit is necessary in case the
relay node is the node that is involved in an accident. Thus,
\begin{equation}
MaxLongBO_d = \frac{R_{tx}}{V_{max}}, \label{eqn:MaxLongBO}
\end{equation}
where, $V_{max}$ is the maximum velocity of the vehicles.

\subsection{Two-Dimensional Scenario}
The forwarding algorithm as described above does not have a
mechanism to select a proper relay in a two-dimensional network,
since all the nodes at equal distance from the sender have equal
probability of becoming a relay. The nodes forwarding message with a
two-dimensional ZOR also face the decision on which transmissions to
accept as implicit acknowledgement.

%
To spread the message throughout the two-dimensional ZOR, the relay
nodes should have a wide angular distance to cover substantially new
regions of the ZOR. Similarly, if a node receives the same message
from relays that cover a major portion of its own coverage area,
there is a high probability that other nodes in its coverage area
would also have received the message and transmission by the node
would be redundant. The ratio of the area of overlap of coverage
area or coverage disk of two or more nodes with respect to their
average coverage area is called \emph{coverage ratio}. Hence, the
angular distance and the coverage ratio of the relays should be
greater than certain thresholds, \emph{angular threshold} and the
\emph{coverage ratio threshold} respectively, to ensure spreading
and flooding of the message.

\begin{figure}[t]
  \center
  \begin{tabular}{p{0.45\textwidth} p{0.45\textwidth}}
    \includegraphics[height=0.45\textwidth,angle=270,clip=true]{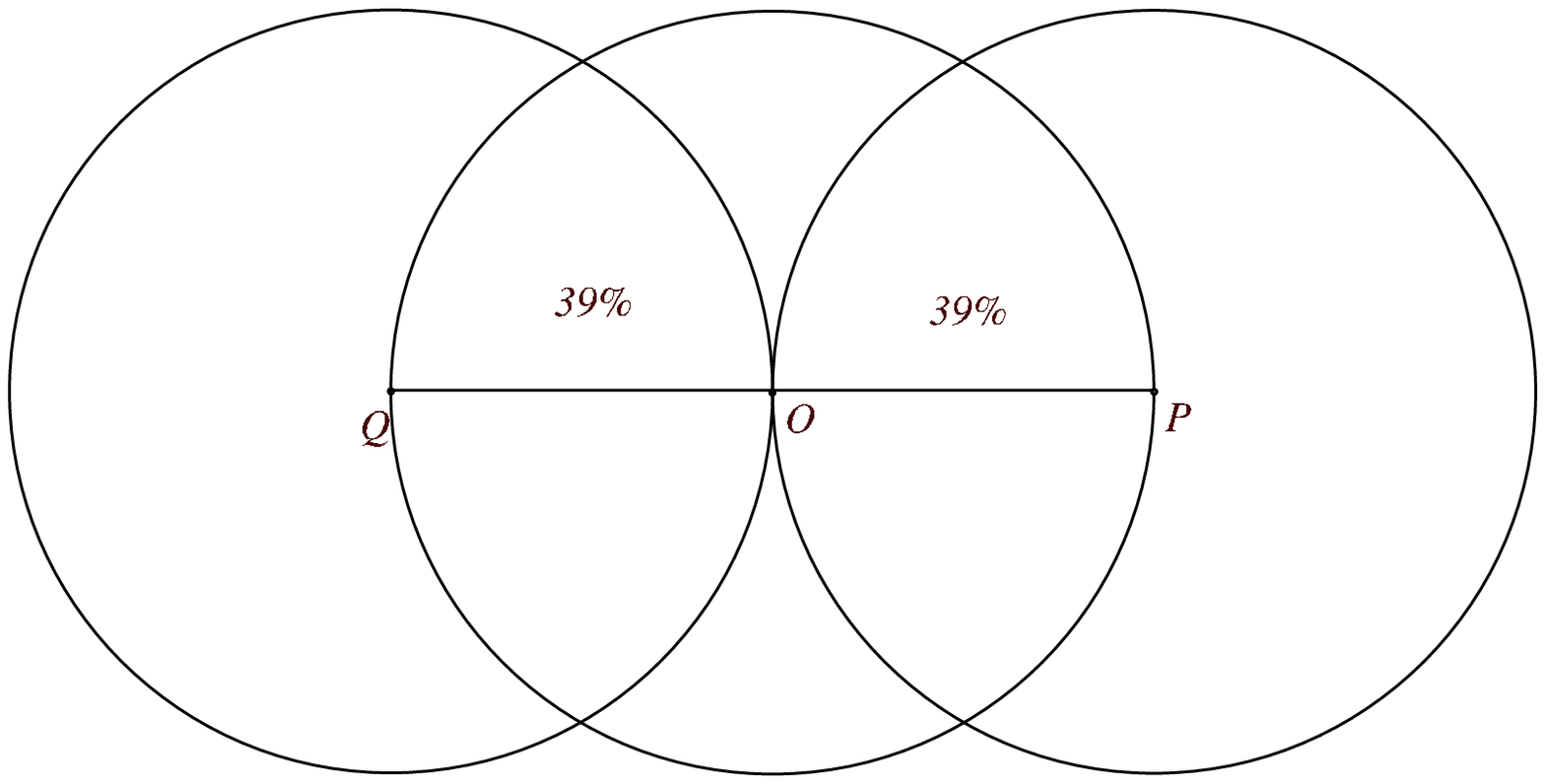} &
    \includegraphics[height=0.45\textwidth,angle=270,clip=true]{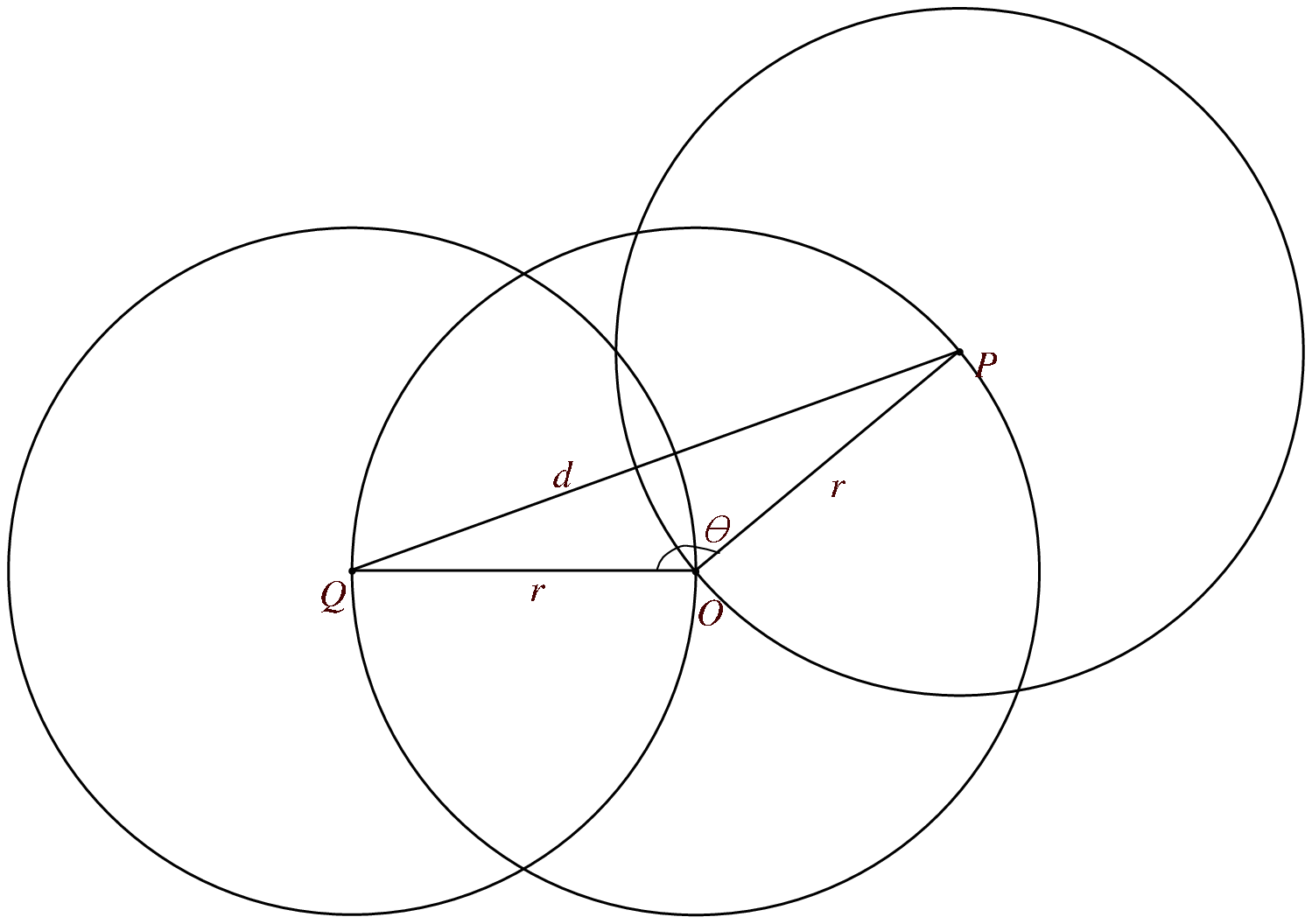} \\
    (a) Two nodes on the edge of center node's transmission range &
    (b) Two nodes forming an angle $\theta$ at the center node \\
  \end{tabular}
  \caption{Two cases of overlap of transmission ranges of two nodes.}
  \label{fig:TxRangeExamples}
\end{figure}

Let us, momentarily, assume a disk model of radio transmission. If
two nodes are at a distance $d$, and have a transmission range
$R_{tx}$, the coverage ratio $CR$ is inversely related to the
distance $d$: it is minimum (zero) for $d \geq 2R_{tx}$, and maximum
(one) for $d = 0$. For two nodes within each other's transmission
range, $CR$ is minimum when $d = R_{tx}$. In~\cite{Joshi:DRG}, for
two nodes within each other's transmission range,
\begin{align}\label{eqn:CRmin}
CR_{min} = \frac{2}{3} - \frac{\sqrt{3}}{2\pi} \approx 0.391.
\end{align}
An ideal scenario for geocast on a straight road is shown in
Fig.~\ref{fig:TxRangeExamples} (a), where nodes $O$ and $P$ relay
the message from $Q$ respectively. From~\eqref{eqn:CRmin}, we know
that the $Q$ and $P$ cover approximately 78\% of node $O$'s coverage
area. If the coverage ratio threshold is higher than 78\%, node $O$
will continue to retransmit the message without any gain in
spreading or flooding of the message. Thus, the upper bound on
coverage ratio threshold $CR_{threshold}$ is:
\begin{equation}\label{eqn:CRthreshold}
CR_{threshold} \leq 0.78.
\end{equation}

The success of the $CR_{threshold}$ criterion depends on a very
accurate estimate of actual transmission range. However, the disk
model assumed here is not very realistic: the actual transmission
range may change with time, and may not be circular in shape. Not
only is the coverage ratio calculation inaccurate, but it also
increases in complexity for multiple nodes. We propose to use angle
based criterion instead by mapping a minimum coverage ratio to an
angle, e.g., coverage ratio of 78\% is mapped to $180^o$. A general
case is shown in Fig.~\ref{fig:TxRangeExamples} (b), where nodes $P$
and $Q$ make an angle $\theta$ at the center node $O$. Let our
desired $CR_{threshold}$ be $x$. What should be the minimum value of
$\theta$ for the minimum coverage ratio to be more than the
threshold $x$. We need to find an angle $\theta$ such that the area
of intersection of disks $P$ and $Q$ should not be more than $(0.78
- x)$, i.e.,
\begin{align} \label{areaPinterQ}
A_{P \cap Q} & \leq (0.78 - x) A_{disk}, \\
A_{P \cap Q} & = 2r^2 \arccos \left(\frac{d}{2r} \right)
    - \frac{d}{2} \sqrt{4r^{2} - d^2},
\end{align}
where $d$ is the distance between nodes $P$ and $Q$.

Without loss of generality, we can assume the disks to be unit
circles, or the transmission range $r$ to be 1. Thus, equation
\eqref{areaPinterQ} becomes,
\begin{align}
2 \arccos \left(\frac{d}{2} \right)
 - \frac{d}{2} \sqrt{4 - d^2} & \leq (0.78 - x) \pi, \label{dLimit}
\end{align}
where $0 < d \leq 2$.

From the Fig.~\ref{fig:TxRangeExamples} (b), the relation between
distance $d$ and angle $\theta$ is:
\begin{align}
\theta & = 2 \arcsin \left(\frac{d}{2r} \right). \label{thetaD}
\end{align}

Thus, from equations \eqref{dLimit} and \eqref{thetaD} we can find a
value of $\theta_{min}$ such that the minimum coverage ratio is
above the $CR_{threshold}$. The calculation of $\theta_{min}$ is
one-time, and significantly reduces the complexity of calculating
coverage ratio by each node by replacing it with simple calculation
of angle between three nodes. Thus, when a node receives a message
from at least two other nodes that make an angle $\theta \geq
\theta_{min}$, the message should be considered to be acknowledged
and spreading in desired direction and all retransmissions of that
message should be canceled since a retransmission will not
significantly add to the coverage.

\subsection{Time Persistence}
We propose a simple yet efficient technique for time persistent
geocast based on periodic rebroadcast approach
of~\cite{MaiStoredGeocast}. Each node sets a persistence timer on
receiving a new geocast message. Upon expiration of the timer, only
those nodes that have not received a transmission of that message
recently, i.e., within recent time threshold for persistence
$T_{R_{p}}$, broadcast the message. To determine the value of
$T_{R_{p}}$, we propose the following formula:
\begin{equation}
T_{R_{p}} = \epsilon \frac{R_{tx}}{V_{max}} + rand(CW_{min},
CW_{max}),
\end{equation}
where, $\epsilon$ is the sensitivity factor, $R_{tx}$ is  the
nominal transmission range, $V_{max}$ is the maximum velocity of the
vehicles, $CW_{min}$ and $CW_{max}$ are the minimum and maximum
collision window respectively, and $rand(a,b)$ is a function that
generates a random number uniformly distributed between $a$ and $b$.

Thus, a new node entering the ZOR can be expected to receive a
transmission of the geocast message $1 / \epsilon$ times before it
reaches one hop distance into the ZOR.

\section{Simulation Environment}\label{sec:SimEnv}
The network simulator SWANS based on the simulation engine
JiST~\cite{SWANS}, along with the STRAW~\cite{Choffnes:TrafficModel}
module is used for evaluating performance. JiST/SWANS is a wireless
simulator, similar to ns-2 and GloMoSim, capable of simulating large
networks. SWANS has radio propagation models including disk model
(i.e. no fading) and Rayleigh fading. STRAW is a mobility model for
vehicles on city streets. STRAW uses a car-following model to model
mobility of vehicles within a road segment. Certain changes, like
implementing lane changing behavior for vehicles in STRAW based on a
model proposed by Kesting et al.~\cite{MOBIL}, and modifying SWANS
to work with geographical addressing, were made.

The performance of DRG is evaluated and compared with a modified
flooding algorithm. The simple flooding algorithm is modified to
restrict the flooding to the ZOR, and to include a collision
avoidance scheme based on random slot backoff. The collision window
and slot size are selected to provide optimum performance in a
typical scenario.

We use a collision warning application as a representative for the
safety applications. In this application, if a vehicle is either
involved in or detects a collision or breakdown, it sends a warning
message to other vehicles. A suitable zone of relevance (ZOR) is
determined by the application. In our simulations, the ZOR is
rectangular with a length $L$ and width $W$, and all the vehicles
within the rectangle are part of the ZOR, regardless of their
direction. The zone of forwarding (ZOF) is defined by adding 15
meters to the bounds of ZOR.

The performance on collision warning application is evaluated on two
scenarios: a straight highway and a city street network. The
performance is evaluated based on three metrics. \emph{Packet
Delivery Ratio (PDR)} is the ratio, as percentage, of the number of
nodes receiving the packet and the number of nodes that were
supposed to receive the packet. When a source generates a new
geocast message for a particular ZOR, a list of nodes belonging to
that ZOR is created and this is used to identify the nodes that are
supposed to receive the geocast message. \emph{End-to-End Delay} is
the time delay between the time a geocast message is sent by an
application at the source node to the time the application running
on receiver node receives the message. \emph{Overhead} is the ratio
of the number of network layer bytes transmitted to the number of
bytes sent by the application layer for a unique message, and is a
measure of efficiency of the routing protocol in reducing redundant
transmissions for restricted flooding based protocol.


\section{Results and Discussion}\label{sec:PerfEval}

We evaluate the performance of the three approaches by varying the
vehicle density on a highway scenario and a city street scenario. We
use a straight highway 10km long and with 3 lanes in each direction.
The maximum speed allowed on the highway is 120km per hour. The
vehicles are placed at a regular distance depending on the density
of the vehicles. The lead vehicle "crashes" three seconds into the
simulation and generates a single collision warning message. The 300
meters wide ZOR starts at the colliding vehicle and extends to 1.5
km behind it. The nominal transmission range is 300 meters.

We use a city scenario with a relatively sparse network and short
transmission ranges to evaluate the performance of the protocols.
The city is a grid of 2km x 2km, streets placed 100 meters apart and
perpendicular to each other. The vehicles are  placed randomly. The
vehicle that sends the collision warning message is always placed at
the center of the grid. The ZOR is a square of $1 km^2$ with the
source node at the center. The default nominal transmission range is
200 meters to account for radio obstacles in a city environment. In
order to show the effect of time-persistent geocast, we set the
time-to-live (TTL) to 15 seconds for DRG. The default value of TTL
for Flooding is 64 hops.

A more detailed discussion of the performance evaluation for
transmission range and the size of the ZOR, and for a traffic
monitoring application, is presented in~\cite{Joshi:DRG}


\begin{figure}[htbp]
\center \subfigure[Highway Scenario]{
  \includegraphics[width=0.47\textwidth]{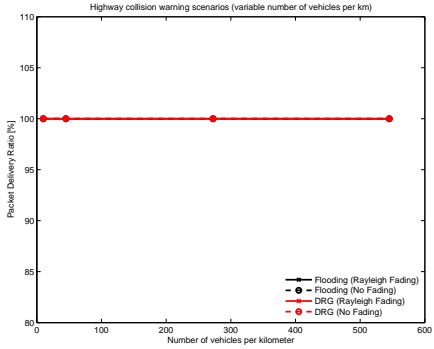}
  \label{fig:hwyPDR}
  }
\hfill \subfigure[City Scenario]{
  \includegraphics[width=0.47\textwidth]{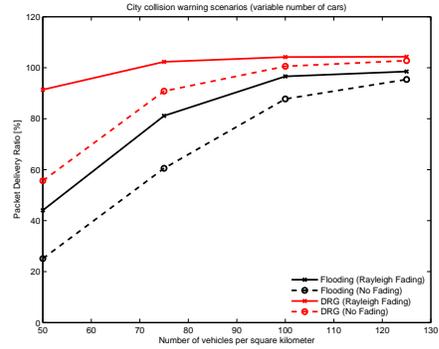}
  \label{fig:cityPDR}
  }
\caption{The average packet delivery ratio as a function of vehicle
density} \label{fig:PDR}
\end{figure}

\subsubsection{PDR}
The PDR for Flooding and DRG in highway scenario, as shown in
Fig.~\ref{fig:hwyPDR}, is 100\% since the network remains connected
even for low vehicle density. In a city scenario,
Fig.~\ref{fig:cityPDR}, the reliability of DRG is much better than
Flooding in a scarce network. This is due to the mechanisms used by
DRG to overcome temporary network fragmentation. Also note that the
PDR is more than 100\% for DRG in certain cases, since the geocast
message is kept alive for 15 seconds by which time new nodes enter
the ZOR and the message is delivered to them.

\begin{figure}[htbp]
\center \subfigure[Highway Scenario]{
  \includegraphics[width=0.47\textwidth]{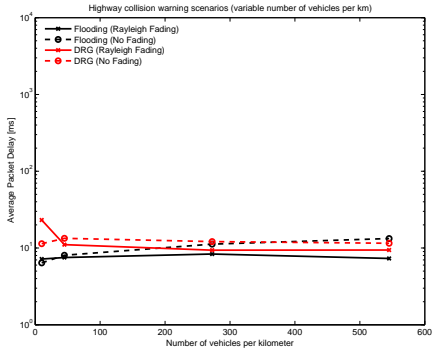}
  \label{fig:hwyDelay}
  }
\hfill \subfigure[City Scenario]{
  \includegraphics[width=0.47\textwidth]{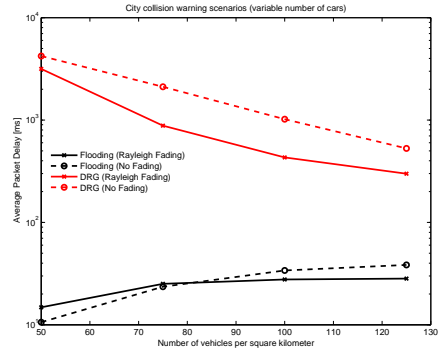}
  \label{fig:cityDelay}
  }
\caption{The average end-to-end delay as a function of vehicle
density} \label{fig:Delay}
\end{figure}

\subsubsection{End-to-End Delay}
The effect of vehicle density on the end-to-end delay is shown in
Fig.~\ref{fig:Delay}. 
With a given coverage area, a higher node density causes more
contentions or collisions for broadcast based protocols like
Flooding, resulting in a higher end-to-end delay. However, the
contention avoidance mechanism introduced for Flooding effectively
reduces the rate of growth in end-to-end delay. The node density
does not significantly affect the end-to-end delay for DRG in a well
connected network.

In a sparse network the DRG delivers to vehicles, once temporarily
separated by network fragmentation, when they enter the coverage
area of a relay. Since, vehicle movements take much longer time than
the time taken by a packet to propagate through a well connected
network, the average end-to-end delay is dominated by the time taken
by vehicles to bridge network fragmentation in a sparse network, as
seen in Fig.~\ref{fig:cityDelay}. However, as the connectivity
improves, the end-to-end delay reduces. The delay still is much
larger than that of simple Flooding, mainly because the geocast
message is kept alive for a long duration, and the message is
delivered to nodes which enter the ZOR even after a long time.

\begin{figure}[htbp]
\center \subfigure[]{
  \includegraphics[width=0.47\textwidth]{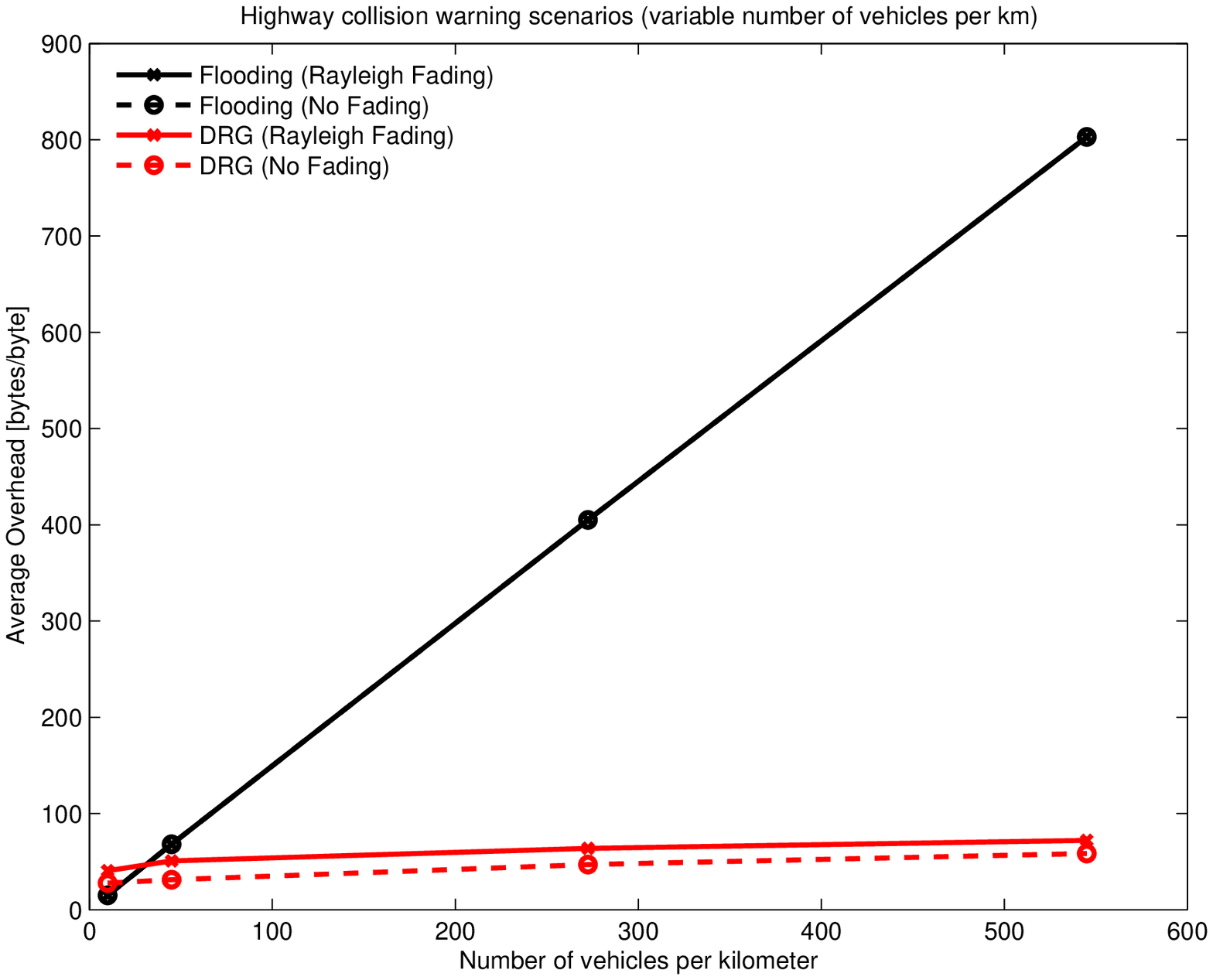}
  \label{fig:hwyOverhead}
  }
\hfill \subfigure[]{
  \includegraphics[width=0.47\textwidth]{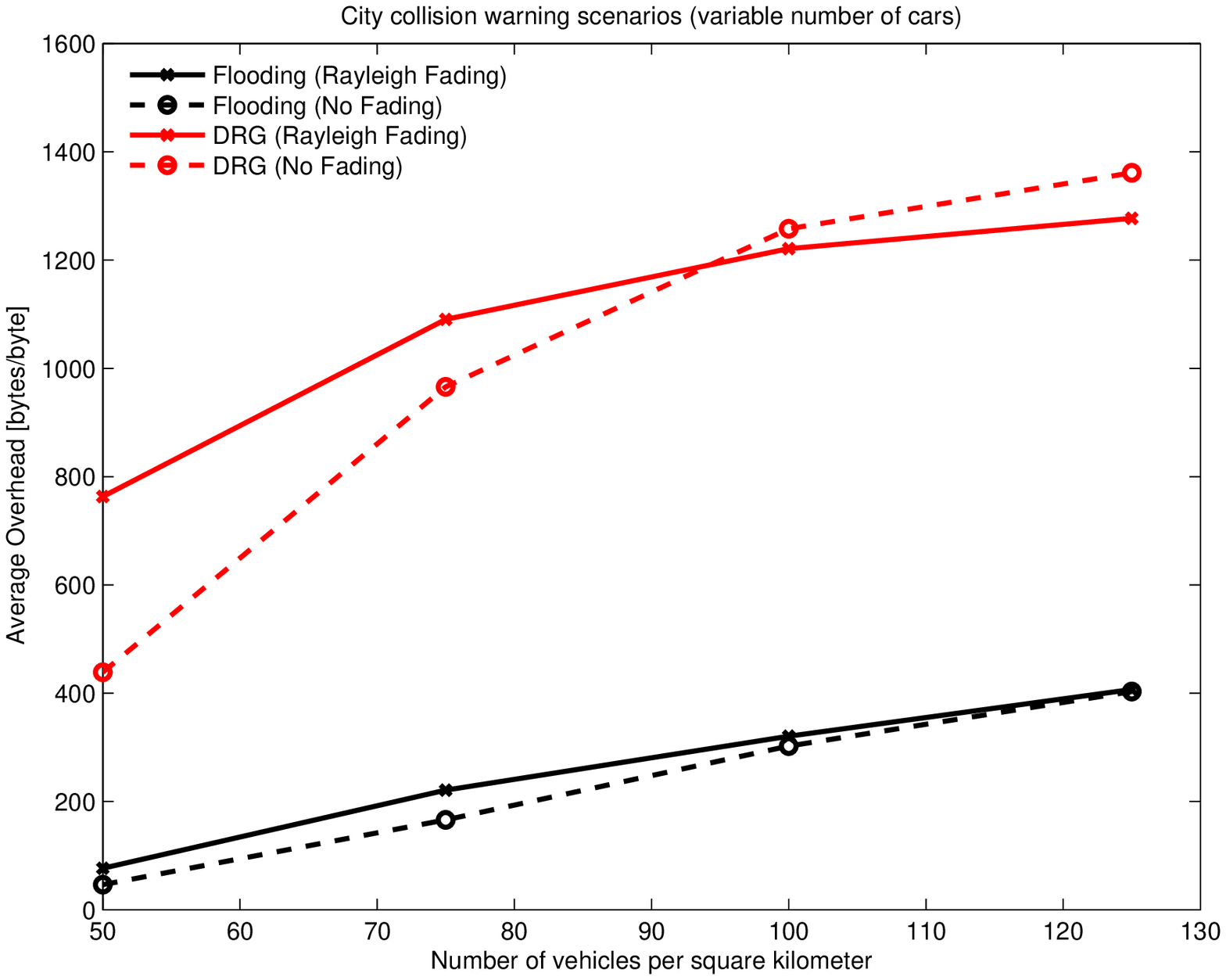}
  \label{fig:cityOverhead}
  }
\caption{The average overhead as a function of vehicle density}
\label{fig:Overhead}
\end{figure}

\subsubsection{Overhead}
The number of transmissions for Flooding is of the order of $O(n)$,
where $n$ is the number of nodes in the ZOR and ZOF. Hence, the
overhead for Flooding increases linearly with the node density. Due
to the distance-based backoff mechanism in DRG, the number of
transmissions for DRG is of the order of $O(k)$, where $k$ is the
number of hops in the ZOR and ZOF.  Thus, the number of transmitting
nodes are not significantly affected by node density. Hence, DRG
scales much better than Flooding in a well connected, dense network
as seen in Fig.~\ref{fig:hwyOverhead}.

The higher PDR for DRG in a fragmented network comes at the cost of
a higher overhead, as seen in Fig.~\ref{fig:cityOverhead}. The
retransmissions to overcome network fragmentation or to keep the
message persist in time add heavily to the overhead. However, the
overhead for DRG grows much slower than that of Flooding or ROVER in
a connected network. Thus, DRG tends to reduce redundancy, when it
is not required to ensure delivery.

\section{Conclusion}\label{sec:Conclusion}
We present algorithms that work in both one-dimensional and
two-dimensional network topology. We have shown through simulations
on various scenarios that while the reliability of DRG is comparable
or even better than that of the highly redundant Flooding, the
overhead is much smaller. The scalability of DRG is also better as
its performance is less sensitive to network size or node density.
However, most importantly, DRG adapts itself to fit network topology
and ensures a high delivery ratio in a sparse and disconnected
network by increasing overhead, while it efficiently delivers the
packets in a well connected and dense network.

\bibliographystyle{splncs}
\bibliography{references,vanet}

\end{document}